\documentclass[twoside]{article}
\usepackage{epsfig.sty}
\newcommand{\beq}{\begin{eqnarray}}
\newcommand{\eeq}{\end{eqnarray}}
\begin{document}

\Large{\bf Gravitational Radiation From Pulsar Creation}
\normalsize
\vspace{5mm}

{\bf Leonard S. Kisslinger$^{1}$, Bijit Singha$^{1}$, Zhou Li-juan$^{2}$

1)Department of Physics, Carnegie Mellon University, Pittsburgh, PA

\hspace{1cm} kissling@andrew.cmu.edu \hspace{1cm} bsingha@andrew.cmu.edu

2)School of Science, Guangxi University of Science and Technology,Guangxi 

\hspace{1cm} zhoulijuan05@hotmail.com

\date{}

\begin{abstract}
  We estimate the gravitational wave amplitude as a function of frequency
produced during the creation of pulsars from the gravitational collapse of
a massive star. The three main quantities needed are the magnitude of the
magnetic field producing pulsar kicks, the temperature which determines
the velocity of the pulsar and the duration time for the gravitational
radiation.
\end{abstract}
\vspace{3mm}


\section{Introduction}

   Shortly after the collapse of a massive star often a neutron star is 
created. Because of the strong magnetic field and high temperature the
neutron star has a large velocity and emits electromagnetic radiation, which
is why it is called a pulsar. The origin of a pulsars velocity is called a
pulsar kick. In research on pulsar kicks\cite{hjk07,kj12} the magnitude of
B, the magnetic field, and T, the temperature of the neutron star at the time
of the pulsar kick, were estimated. We use these parameters in the
present estimate of gravitational wave creation from pulsar creation, for
which there are no published previous estimates.

In a study of gravitaional waves (GW) generated by magnetic
fields\cite{kks10} it was found that the GW created during the EWPT could be
detected by Lisa Interfer0meter Space Antena (LISA) while the GW created
during the QCDPT cannnot be detected by LISA.
There is also an article\cite{kk15} on the circular polarization of
GW created during the EWPT and QCDPT.

  Previous research on gravitational radiation from neutron stars\cite{mp10,
lbm13} was based in part on gravitational radiation from cosmological 
turbulance\cite{kmk02}.

There have been studies of gravity waves generated by cosmological phase
transitions, the Electroweak Phase Transition (EWPT) and the Quantum
Chromodynamic Phase Transition (QCDPT). In a study of gravity waves generated
by magnetic fields\cite{kks10} it was found that the gravitational waves
generated during the EWPT might be detected by the Laser Interoferometer
Space Antenna (LISA) while gravitational waves generated during the QCDPI
canot be detected by LISA.

With current theory the EWPT and QCDPT are first order phase thansitions,
so bubbles of the new universe form within the older universe during the
phase transition. A study of gravitational waves created by bubble collisions
during the EWPT and QCDPT estimated the degree of circular polarization
of the gravitational waves\cite{kk15}.
 
The formalism used Refs\cite{kks10,kk15} is not suitable for estimates of the 
creation of gravitational radiation from pulsar creation. However, the research
in Refs\cite{mp10,lbm13,kmk02} was based in part on gravitational radiation
from cosmological turbulance which we use in our present research.

\newpage

\section{Theory of gravitational radiation from magnetic fields}

Einstein's General Theory of Relativity is based on the equation (with no 
cosmological constant)
\beq
\label{Ein-genrel}
    {\cal{R}}^{\alpha \beta}- \frac{1}{2}{\cal{R}}g^{\alpha \beta} &=& 8 \pi  G 
T^{\alpha \beta} \; ,
\eeq
 
where $\mathcal{R}^{\alpha \beta}, \mathcal{R}$ the Ricci tensor, Ricci scalar,
$g^{\alpha \beta}$ is the metric tensor, $T^{\alpha \beta}$ is the energy-momentum
tensor and G is Newton's gravitational constant. 

  In Ref.\cite{kmk02} $T_{i j}(x)$, with $i,j$ =1,2,3 is 
\beq
\label{Tij(x)}
  T_{i j}(x)&=& w u_i(x)u_j(x) \; ,
\eeq 
with $w=\rho$, the mass density, and $u$ the velocity of the material emitting
the gravitational radiation. In Fourier space at time $t$
\beq
\label{Tij(k)}
  T_{i j}(k,t)&=&\frac{V}{(2\pi)^3} w \int dq  u_i(q,t)u_j(k-q,t) \; ,
\eeq
where $V$ is a volume factor. 

The energy density in gravitational radiation scales like $a^{-4}$, where
$a$ when the temperature of the universe was $T$, with $a_o$ the scale factor
now, is
\beq
\label{a-T}
 \frac{a}{a_o}&=&8.0\times 10^{-16} (100/g)^{1/3}(100 {\rm GeV}/T) \; ,
\eeq 
where $g$ is the number of degrees of freedom at temperature $T$.

The energy momentum tensor from a magnetic field $\vec{B}$ is
\beq
\label{T-B}
 T^{(B)}_{i j}(k,t)&=&\frac{V}{(2\pi)^3 4\pi}  \int dq \large [ B_i(q,t)
B_j(k-q,t)-\delta_{i j}  B_l(q,t)B_l(k-q,t) \large ] \; ,
\eeq
where $t$ is the time the  magnetic field $B$ is generated.

  The amplitude of the gravitational wave created by magnetic fields, $h_c^B$,
as a function of $f$, the frequency of the gravitational radiation
today, and $f_B=\tau_B^{-1}$, where $\tau_B$ is the duration time of the 
source of the $B$ field from Ref\cite{kmk02} Eq(75) is
\beq
\label{hcB}
   h_c^B(f,\eta_o)&=&\frac{3^{3/2}}{2^{5/3}\pi^{2/3}}\frac{G w 
\epsilon^{2/3}f_B^{-1/3}a^{14/3}}{H_o\sqrt{\Omega_{rad}}} 
f^{-4/3}\xi(f\eta_o,f\eta_{end})\\
  \epsilon&\simeq& \frac{27}{8} k_D^4 \nu^3 \nonumber \\
 \xi(f\eta_o,f\eta_{end})&\equiv& \xi(f)= \int_{\eta_o}^{\eta_{end}} d \eta'
\frac{sin(f\eta_{end}-f\eta'))}{\eta'} \nonumber \\
  H_o\sqrt{\Omega_{rad}}&=& a_o/\eta_{end} \nonumber \\
  w &=& B^2/(4 \pi u_B^2) \nonumber  \; ,
\eeq
\newpage
\noindent
with $k_D$ the wave number corresponding to the smallest-scale motions,
$\nu$ is the kinematic viscosity of the source, $\eta_o,\eta_{end}$ are the
conformal time at the beginning and end of the process creating the 
gravitational wave, and $u_B$ is the velocity associated with the creation of 
the $B$ field.

  Therefore, to determine the gravitational wave amplitude produced by
magnetic field generation one needs the magnitude of the magnetic field, B, the 
temperature, T, and the duration time, $\tau_B$. These quantities are also 
needed for our estimate of the gravitational wave amplitude produced by pulsar 
creation, which is reviewed in the next section.

\section{Pulsar Kicks With Magnetic Field B and  Temperature T}

  Neutron stars, often called pulsars, are created by the gravitational 
collapse of a massive star. The pulsars move with much greater velocities than 
other stars in our galaxy, called the pulsar kick. Neutrinos produced by
the URCA process dominate the emission of energy during the first 10-20 seconds
after the collapse of a heavy star\cite{hjk07}, with the URCA process which
dominates neutrino prodution in neutron stars\cite{bw65}.
\beq
\label{URCA}
  n+n &\rightarrow& n+p+e^- +\bar{\nu}^e
\ ,
\eeq
with $\bar{\nu}^e$ an anti-electron neutrino. The electrons are in Landau
levels due to the strong magnetic field B. With a magnetic field $B \simeq
10^{16}$ Gauss, which is created before the URCA process\cite{hjk07}, the 
momentum given to the pulsar with  the duration time $\tau_T=1/f_B \simeq 10$s,
\beq
\label{pns}
   p_{ns} &\simeq& 0.43 \times 10^{27} (T/10^9 K)^7 (R_{ns}^3-R_{\nu}^3)
\rm{gm\;cm\;s^{-1}} \; ,
\eeq
with $R_{ns},R_{\nu}$ rhe radius of the protoneutron star and the radius of the
neutrinosphere. 

In Ref.\cite{hjk07}  $R_{\nu}$ was estimated using a standard value for $R_{ns}$,
with the result
\beq
\label{RnsRnu}
     R_{ns} &=& 10 {\rm \;\;km} \nonumber \\
     R_{\nu} &\simeq& 9.96 {\rm \;\;km} \; .
\eeq

From Eqs(\ref{pns},\ref{RnsRnu})
\beq
\label{pns2}
  p_{ns} &\simeq& 5.14 \times 10^{-4} (T/10^{10})^7  = M_{ns}v_{ns} \; ,
\eeq
where $M_{ns}$ is the mass of the neutron star and $v_{ns}$ is the velocity
of the neutron star. 

 Using $M_{ns}=2\times10^{33}$ gm $\simeq$ the mass of the sun,
and including a factor of 0.4, from Eq(\ref{pns2} the velocity of the neutron 
star $v_{ns}$ is
\beq
\label{vns}
  v_{ns}&\simeq& 1.03 \times 10^{-4} (T/10^{10})^7 {\rm \;\;km/s} \; .
\eeq
\newpage

From Eq(\ref{vns}) one obtains the pulsar velocity as a function of temperature,
$v_{ns}(T)$, as shown in Figure 1 below.
\vspace{2cm}

\begin{figure}[ht]
\epsfig{file=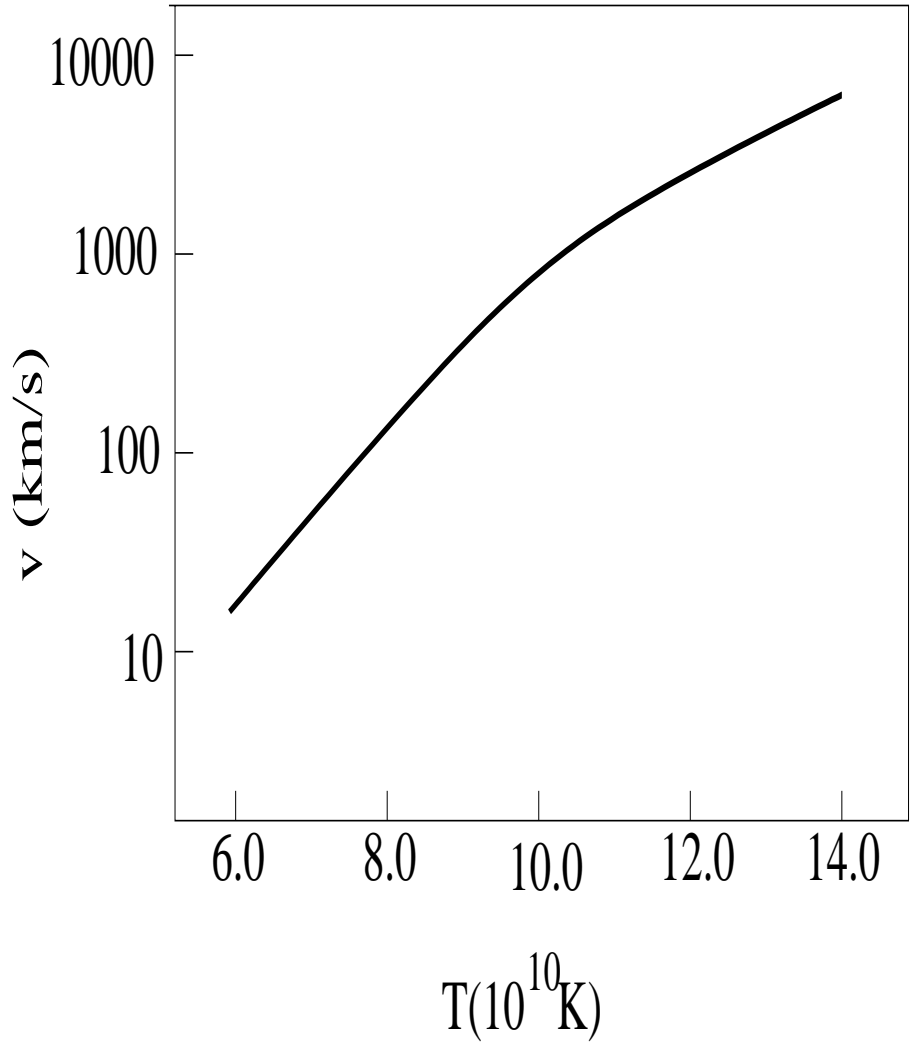,height=14cm,width=12cm}
\caption{The pulsar velocity as a function of T}
{\label{Fig.1}}
\end{figure}
\newpage

\section{Gravitational Radiation From Pulsar Creation}

  From sections Theory of gravitational radiation from magnetic fields and
Pulsar Kicks With Magnetic Field B and  Temperature T, we now estimate
the gravitational wave amplitude as a function of the linear frequency, 
$h_C(f)$, from the creation of pulsars. 
\vspace{5mm}

Note: In order to calculate $h_C(f)$ and plot it as a function of $f$ as in
Ref\cite{kks10} we must determine the quantities $\epsilon$, $\xi(f\eta_o,
f\eta_{end})$, $ H_o\sqrt{\Omega_{rad}}$. and $w$ in Eq(\ref{hcB}), using
Refs\cite{hjk07},\cite{kj12} and other articles related to pulsar creation.

The velocity associated with the creation of the B field, $u_B$, has been
estimated by measuring the radial velocities of OBN (enhanced nitrogen)
stars\cite{lmm88} for a period of 27 days, with the result
\beq
\label{uB}
     u_B &\simeq& 90 {\rm \;km/s} \; .
\eeq

Fron Eq(\ref{uB}) and using $B \simeq 10^{16}$ Gauss or, 
$B \simeq 10^{19}  M_W^2$, with $M_W\simeq$ 80 GeV
\beq
\label{w}
  w &=& B^2/(4 \pi u_B^2) \simeq 6.4 \times 10^{36} {\rm GeV^2 s^2/km^2}
\; .
\eeq

  From Ref\cite{hjk07} the duration time $\tau_B \simeq$ 1 s, so
\beq
\label{fB}
 f_B &\simeq& 1.0 {\rm \;s^{-1}} \; .
\eeq 
 From Ref\cite{kmk02} Eq(14) and Eq(\ref{hcB}) 
\beq
\label{kd4} 
  k_D^4 &=& \frac{8 \kappa \rho_{\rm vac}}{27 \nu^3 \tau_s w} \\
 \epsilon &\simeq& \frac{\kappa  \rho_{\rm vac}}{\tau_s w} \nonumber \; ,
\eeq
with the source time interval  $\tau_s \simeq$ 1.0 s and the efficiency 
$\kappa \simeq$ 1.0. From Eqs(\ref{hcB},\ref{kd4})
\beq
\label{hcB2}
   h_c^B(f,\eta_o)&=&\frac{3^{3/2}}{2^{5/3}\pi^{2/3}}\frac{G w^{1/3} 
(\rho_{\rm vac}/\tau_s)^{2/3} f_B^{-1/3}a^{14/3}}{a_o/\eta_{end}} 
f^{-4/3}\xi(f\eta_o,f\eta_{end}) \nonumber \\
&=& 0.76 G w^{1/3}(\rho_{\rm vac}/\tau_s)^{2/3} 
f_B^{-1/3}(a/a_o)^{14/3}a_o^{11/3}\eta_{end}f^{-4/3}\xi(f) \; ,
\eeq
\noindent
where $\eta_o,\eta_{end}$=(1s,2s) are the conformal time at the beginning and 
end of the process creating the gravitational wave.
\newpage

  Defining $A$ by
\beq
\label{hcB2A}
  h_c^B(f,\eta_o)&=& A f^{-4/3}\xi(f) \; ,
\eeq
using $c=c\not h=1.0$, Newton's gravitational constant $G=6.71\times 10^{-39}
 \rm{GeV^2} $, the temperature\cite{hjk07} $T=10^{10}$ K so 
$(a/a_o)^{14/3}\simeq 105.0$, free energy
density $\rho_{\rm vac}=9.9\times 10^{-30} \rm{gm}/(\rm{cm}^3)$, and the other
parameters in Eq({\ref{hcB2}) given above, one finds
\beq
\label{A}
  A &\simeq& 5.27 \times 10^{-21} {\rm \;\;cm} \; .
\eeq

Our results for the gravitational wave amplitude produced via pulsar
creation are shown in Figure 2. 

\vspace{2cm}

\begin{figure}[ht]
\begin{center}
\epsfig{file=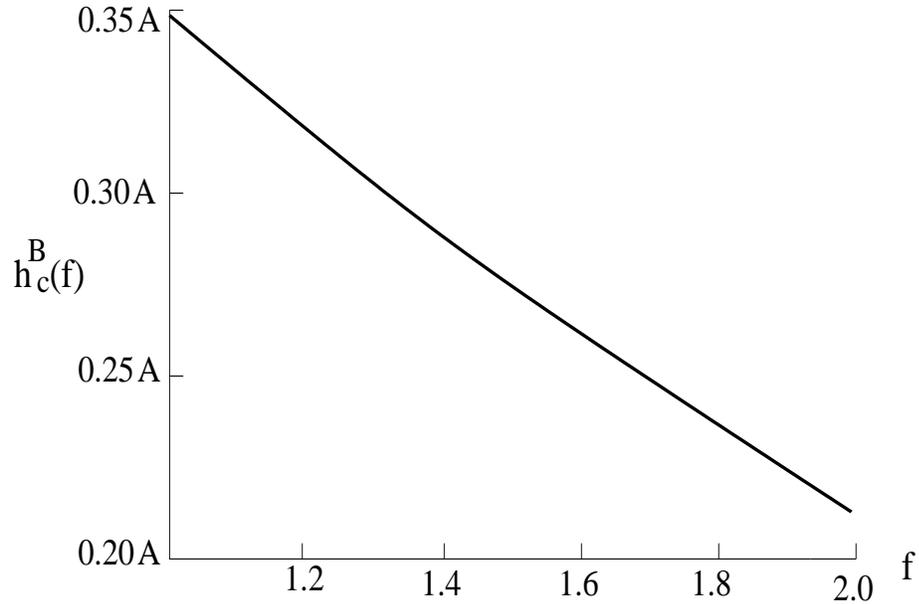,height=8cm,width=12cm}
\caption{h$^B_c$  (f)=A$f^{-4/3}\xi(f)$}
\label{Figure 2}
\end{center}
\end{figure}

\section{Conclusions}

  Our conclusion, based on our results shown in Figure 2, is that the
gravitational wave amplitude produced via pulsar creation is smaller than
that produced via the Cosmological Electroweak Phase Transition\cite{kks10}
and is too small to be measured by the Lisa Interferometer Space Antenna 
(LISA)\cite{cornish01} or any other gravitational wave detector in the near 
future.
\newpage
\Large

{\bf Acknowledgements:}
\vspace{2mm}

\normalsize
{\bf Authors thank Prof. Tina Kahniashvili for helpful suggestions. Authors
Leonard S. Kisslinger and Bijit Singha thank the CMU Physics Depart for 
support. 

Author Zhou Li-juan acknowledges the support
in part by the National Natural Science Foundation of China (11365002),\\
the National Natural Science Foundation of China (11865005),\\ 
Guangxi Natural Science Foundation (2015GXNSFAA139012).}

\end{document}